# Discovery of Bayes' Table at Tunbridge Wells

David C. Schneider
Department of Ocean Sciences
Memorial University, St. John's NL A1C5S7 Canada

Roy Thompson
5 College Drive, Tunbridge Wells, Kent, TN2 3PN UK

In 1755 Thomas Bayes expressed an interest in the problem of combining repeated measurements of the location of a star. Bayes described a tandem set-up of a ball thrown on a table, followed by repeated throws of a second ball. Bayes' table has long been taken as a billiard table, for which there is no evidence. We report the discovery of Bayes' table, a bowling green located half a km uphill (SE) from the meeting house where Bayes served as minister for two decades. Bayes' drawing shows a rectangular space marked off in yards, which allows calculation of an interval measurement of uncertainty. The Bayes rule interval from 2.5% to 97.5% is from 0.56 – 0.42 = 0.12 perches equivalent to 0.61 m. The discovery of Bayes' table establishes the physical basis for Bayes' symmetrical probability model, a fixed parameter binomial ($\theta = 0.5$). The discovery establishes Bayes as the founder of statistical science, defined as the application of mathematics to scientific measurement.

Keywords: Thomas Bayes; 18th century science; Bayes' table; Bayes' rule

## INTRODUCTION

The application of mathematical statistics to scientific data began in the 18$^{th}$ century with the problem of combining measurements of the location of a star. In 1755 Simpson showed that a mean based on 6 observations is better than a single observation.[1] In a comment on Simpson's publication Bayes noted that the statement was "only true where the chances for the errors of the same magnitude, in excess or defect, are upon an average nearly equal."[2] Bayes never published on the topic. After Bayes' death his *Method of Calculating the Exact Probability of All Conclusions founded on Induction* was read to the Royal society by Richard Price, who added an introduction, added new material, and changed the title[3] to *An Essay Towards Solving a Problem in the Doctrine of* Chances.[4] In the essay Bayes developed a series of 10 propositions leading to a rule for calculating the probability of an event lying between two degrees of probability around the location of a ball thrown at random. Bayes stated the conditional probability of Event 2 given Event 1 (Prop 3) and then Event 1 given Event 2 (Prop 5). Bayes never produced a theorem that combines the two propositions. Instead, Bayes drew a figure showing the first event--a ball thrown on a table, along with the distribution of the second event—a second ball thrown repeatedly on the table. Bayes used the figure to develop a rule for calculating a probability interval around the first ball.

## BAYES' TABLE

To develop the rule Bayes described a "square table made level." Bayes then described a tandem chance set-up.

"I suppose that the ball W shall be 1st thrown, and that through the point where it rests a line shall be drawn parallel to AD, and meeting CD and AB in s and o; and that afterwards the ball O shall be thrown p+q or n times, and that its resting between AD and os after a single throw be called the happening of the event M in a single trial."

Bayes drawing (Figure 1) shows a rectangular area ABCD, divided into two parts by the line os. It also shows a symmetrical binomial distribution with boundaries at points A and B, where successes p in n trials are to the right of line os. 19th century treatments of Bayes' essay either ignore the table (random draws from a bag[5,6] or urn[7]) or describe random throws on a billiard table.[8] This interpretation has persisted ever since.[9,10,11] Twentieth century interpretations usually refer to a billiard table although a few are careful to either omit billiards[12,13] or are careful to note the absence of any mention of billiards in Bayes' essay.[14,15] Bellhouse was unable to find any clear connection between Bayes and a billiard table.[16] He noted that balls were swatted with a bat (18th century) or later poked with the tail (queue) end of a bat, not thrown.

To those who have done an end of curling (DCS) or lawn bowls (RT) Bayes' drawing is that of the throw of a rock (curling) or a bowl on a rink. Curling existed in the 18$^{th}$ century in Scotland where Bayes attended university.[17] The connection to curling, while possible, is as speculative as that of a billiard table. Turning to evidence, a bowling green existed on Mount Sion at the end of the 17$^{th}$ century.[18] That green no longer exists, but emerges by measuring the distance and direction from Bayes' meeting house on Mt Sion Road to the green on Bowra's 1738 map showing the green and the meeting house (Figure 2).[19] The map distance is 19.7 perches (98.8 m) to the green, which is now occupied by a row of houses that align with the green at the same noticeable angle to Mount Sion Road.

Bayes' drawing (Figure 1) shows several details that otherwise defy explanation. The figure contains parallel lines of 55 equally spaced dots. This scales Bayes drawing to the 10 perch (55 yard) green on the Bowra map; it puts a scale of one perch on the binomial distribution bounded at A and B in Bayes' drawing. The labelling of the balls appears idiosyncratic—one ball labelled W, the other shown as a typeset circle o in the original 1764 publication. In lawn bowling the weighted ball is called a Wood; the other ball is smaller, perfectly round, and unweighted. The mnemonic labelling reverses the usual order of play where the unbiased ball o is thrown first, becoming the target of throws of the Woods. A Wood rolls with a bias, coming to rest with a random component along a line perpendicular to the length of the rink. Rolled first, the Wood becomes the analog of a randomly located star. The second ball, aimed at the first, passes to the left or right of the Wood. Bayes' drawing shows an idealized bowling green measured in yards, as the analog of unbiased and randomly varying measurements aimed at a star.

## BAYES' RULE

Bowra's map showing both bowling green and meeting house connects a finite space measured in perches and marked off in yards to Bayes drawing. This allows the first Bayesian calculation of uncertainty "between any two degrees of probability" and hence between any two points in units of length on the table. We undertook "Bayesian bowling" defined as throwing the Jack, ball o, repeatedly at a Wood resting at a random location on the current green in Tunbridge Wells. Using Bayesian scoring as passing to the right of the Wood by a skilled bowler, we recorded 73 passes in 156 throws, for a proportion of $p/n$ = 0.468. The symmetrical Bayesian interval ("probability that I am right") taken as the interval from 2.5% to 97.5% was $[Pr(p_{upper}) \leq 0.975] - [Pr(p_{lower}) < 0.025] = [Pr(85) \leq 0.975] - [Pr(66) < 0.025] = 95\%$, calculated from the cumulative inverse Binomial ($\theta$=0.5) distribution. The corresponding distance was (85-66)/156 = 0.12 perches, or 0.61 m, taking the distance from A to B in Bayes' drawing as the width of a rink, 1 perch (5.0292 m) on Bowra's map.  The accuracy of this estimate rests on two assumptions: symmetrical placement of the measurement scale (line AB in Bayes' drawing, taken as 1 perch) and symmetrical errors of throws relative to the target.  Turning to fit of the data to the probability model, a 20th century concept, the likelihood ratio relative to a symmetrical binomial was $LR = (73/78)^{73}(83/78)^{83} = 1.4$. The observed proportion $p/n = 0.468$ is just as likely (LR < 5) as 0.5. The non-Pearsonian chi-square fit was $G^2 = 2ln(LR) = 0.64$, for which the probability (0.42) is far too large to reject the hypothesis of no difference between data and model.

## CONTEXT AND CRITIQUES

The discovery of Bayes' table replaces a peculiar and unnatural use of a billiard table with the natural use of bowling green. It replaces random throws (essentially an urn model for blind throws) with targeted throws.  It displaces a concept unknown to Bayes-- a uniform prior probability--with a distribution known to Bayes, a symmetrical binomial. Targeted measurement on a bowling green is consistent with Bayes' description "there shall be the same probability that it [any ball] rests upon any one equal part of the plane as another." It uses a tandem setup for targeted measurement instead of an urn model based on random throws.

The fixed parameter interpretation of Bayes' rule is consistent with the fixed parameter interpretation of Bayes by Fisher and Neyman.[20] It stands in contrast to pervasive and established testimony for random draws or throws, which has remained in place to the present in academic publications,[21] in some[22] but not all[23] texts, and in the popular literature.[24]  The first published interpretation of Bayes' physical model (blind draws from an urn) was by Cournot, in France, where lawn bowling was little known. The chances of a wrong interpretation by Cournot were thus more than negligible. Keynes used his inverse probability theorem to demonstrate the dependence of subsequent testimony on initial testimony.[25]

Would a non-conformist minister have any experience with lawn bowling? Regardless of whether Bayes ever bowled, the rules for bowling were widely known in England in Bayes' time.  Shakespeare's plays contain several references to bowling.  The best known -- aye there's the rub -- refers to the uncertainty in outcome due to the 'rub' of the green on the roll of a ball. Bowling greens were places for social interaction among the wealthy, such as Bayes, regardless of whether they bowled.[26] Bayes would have been familiar with probabilistic treatments of bowling from de Moivre's analysis of betting odds in games of chance.[27]

It has become customary to comment on why Bayes did not publish his findings. Explanations based on random throws (*e.g.* Good) become moot in light of targeted throws of the second ball.[28]  Remaining explanations include Bayes' declining health in 1755,[29] reluctance to publish a rule based on a questionable axiom,[30] inability to establish the truth of Bayes' Proposition 5.[31] Another is lack of an accepted standard of reference that is 'good enough'.[32] The accepted standard for Type I error on a null hypothesis test became 5% in the 20[th] century. Standards for likelihood ratios are still in a state of flux.[33] The discovery of Bayes' table with targeted throws raises several explanations.  First is validity of the model. The 20th century goodness of fit test, as above, addresses this. Another is the loss of information when scoring errors by sign (left or right) rather than by distance from the target. The solution known to us, the normal distribution, was proposed five decades after Bayes' death by Gauss[34] and then established on the basis of the central limit theorem by Laplace.[35] Yet another is the limitation of the binomial distribution to a fixed interval (A to B in Bayes' drawing).  The estimate of the uncertainty interval depends on the width of the interval, which was taken as 1 perch in the exemplary calculation from Bowra's map.  In the case of astrometry, interval width is addressed by the field of view of the telescope.  A related limitation is centering the field of view on the star.  Bayes' symmetrical error premise addresses this assumption.

BAYES' CONTRIBUTION

We conclude that Bayes' probability model is that of targeted measurement of a randomly located object at rest, for which Bayes used a tandem setup with symmetrical binomial distribution.  Prior to Bayes' publication mathematical treatments of chance phenomena were directed at expected values in games of chance, from Pascal to de Moivre.[36]  Bayes did not write the inverse theorem named for him.[37] Instead, his major contribution was to apply conditional probability to the problem of variable measurements in science.[38] The discovery of Bayes' table establishes Bayes as the founder of statistical science, defined as mathematical treatment of scientific measurement.

it on Mount Sion and another up the hill Called Mount Ephraim...so the Gentlemen Bowle, the Ladies dance or walke in the green in the afternoons"

Andrew I. Dale, Bayes or Laplace? An examination of the origin and early applications of Bayes' theorem. *Arch. Hist. Exact Sciences* **27**, 23–47 (1982); Dale, *Most Honourable Remembrance: The Life and Work of Thomas Bayes*, (Springer 2003). Dale gives an exhaustive list of the many renderings of 'Bayes Theorem.' The remarkable variety is symptomatic of the absence of any statement of the theorem by Bayes.

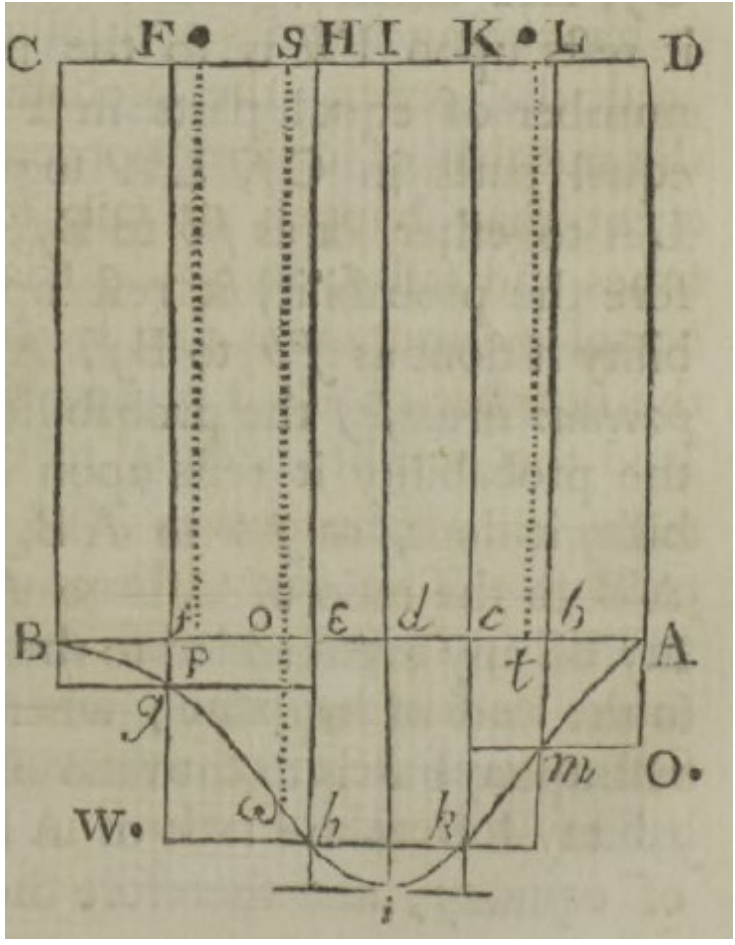

Figure 1. Bayes' drawing of a square table. Taken from unreprinted version of Bayes (1763). Note the labelling of the first ball W and the second ball O.

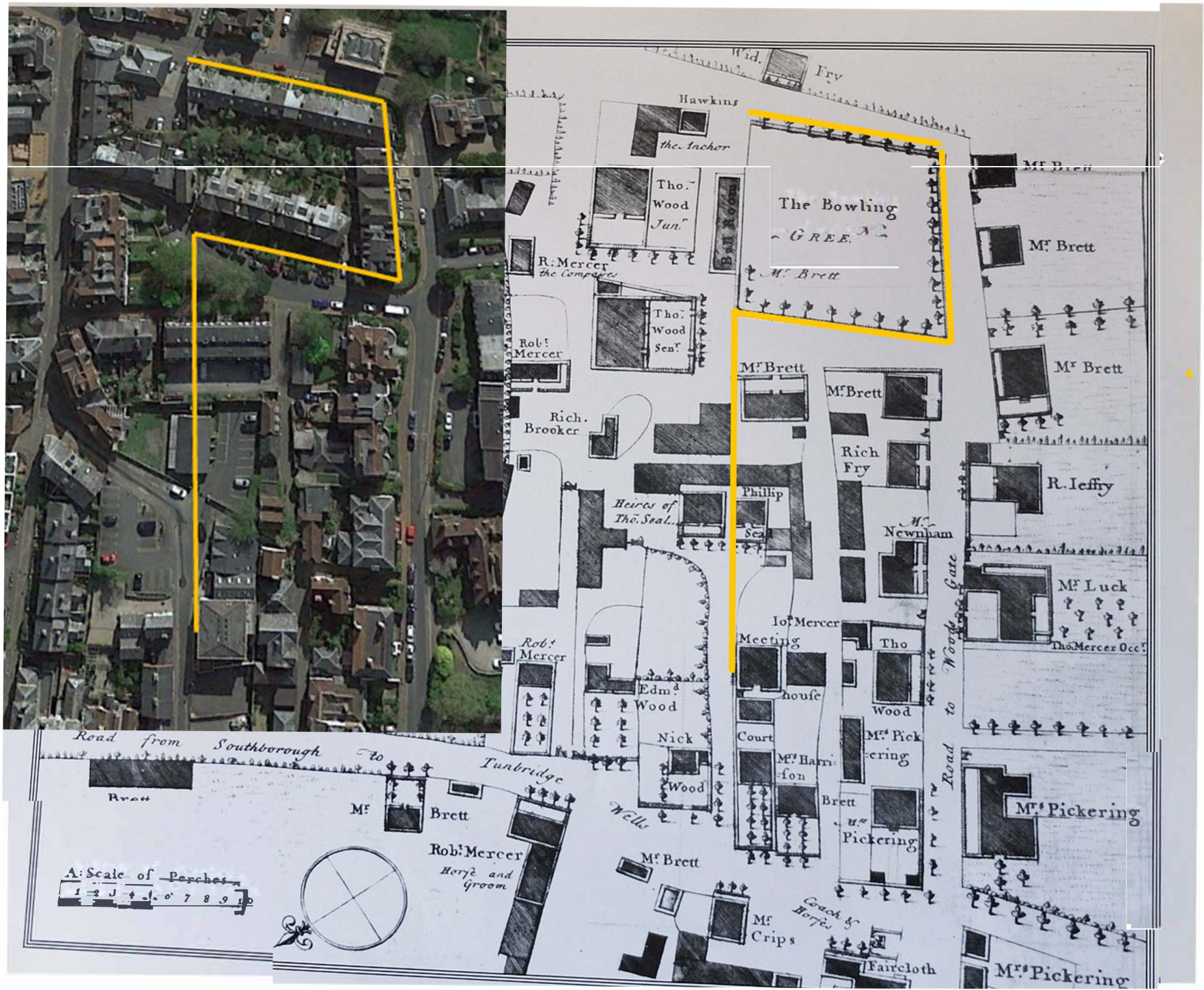

Figure 2. Aerial photograph aligned with map of Mount Sion (Bowra 1738) showing scale of perches and distance from Meeting House to Bowling Green. Note the direction of north on the Bowra map.